\documentstyle[prabib,aps]{revtex}
\begin{document}

\title{Non-differentiable Degrees of Freedom : Fluctuating
Metric Signature}
\author{V. Dzhunushaliev
\thanks{E-Mail Address : dzhun@freenet.bishkek.su}}
\address{Universit\"at Potsdam, Institute f\"ur Mathematik,
14469, Potsdam, Germany \\
and Theor. Phys. Dept. KSNU, 720024, Bishkek, Kyrgyzstan}
\author{D. Singleton
\thanks{E-Mail Address : dougs@csufresno.edu}}
\address{Dept. of Phys. CSU Fresno, 2345 East San Ramon Ave.
M/S 37 Fresno, CA 93740-8031, USA}

\maketitle

\begin{abstract}
In this article we investigate the metric signature
as a non-differentiable ({\it i.e.} discrete as
opposed to continuous) degree of freedom.
The specific model is a vacuum 7D Universe on
the principal bundle with an SU(2) structural group.
An analytical solution is found which to
a 4D observer appears as a flat Universe
with a fluctuating metric signature, and
frozen extra dimensions with an SU(2) instanton 
gauge field. A piece of this solution  with linear
size of the Planck length ($\approx l_{Pl}$)
can be considered as seeding the
quantum birth of a regular Universe. A boundary
of this piece can initiate the formation of a Lorentzian
Universe filled with the gauge fields and in which the
extra dimensions have been ``frozen''.
\end{abstract}

\section{Introduction}

Ref. \cite{dzh99nd} presented a model for the quantum birth
of a 5D Universe from ``Nothing'' via a metric with fluctuating
signature. In this scenario the $5^{th}$ dimension
is associated with a U(1) gauge group ({\it i.e.} the 5D spacetime
is the total space of the principal bundle with U(1) as the 
structural group). In this case the electromagnetic gauge 
field appeared as the non-diagonal components $G_{5\mu}$
($\mu=0,1,2,3$) of the 5D metric. The basic idea in this model was
that the signature of the metric, $\eta_{\bar A\bar B}$, 
($\bar A, \bar B=0,1,2,3,5,6,\cdots$ are viel-bein indices) 
was an independent degree of freedom from the viel-bein 
\begin{equation}
ds^2_{(MD)} = \eta_{\bar A\bar B} 
\left (h^{\bar A}_C dx^C \right )
\left (h^{\bar B}_D dx^D \right )
\label{int1}
\end{equation}
$C,D$ are the multidimensional (MD) coordinate indices,
$x^A$ are the coordinates on the total space of the principal bundle 
with a structural group $\cal G$. The metric can be rewritten
\begin{equation}
ds^2_{(MD)} = \eta_{\bar a\bar b} 
\left ( 
h^{\bar a}_c dx^c + h^{\bar a}_\mu dx^\mu
\right )
\left ( 
h^{\bar b}_c dx^c + h^{\bar b}_\mu dx^\mu
\right ) + 
\eta_{\bar \mu\bar \nu} 
\left ( 
h^{\bar \mu}_\alpha dx^\alpha 
\right )
\left ( 
h^{\bar \nu}_\beta dx^\beta 
\right )
\label{int2}
\end{equation}
$\bar a,\bar b$ are the viel-bein indices for the fibre 
of the principal bundle, and $c,d$ are the coordinate indices on 
the fibre; $\bar\mu ,\bar\nu$ and $\alpha,\beta$ play the 
same role for the 4D base of the principal bundle. From Eqs. (\ref{int1}), 
(\ref{int2}) we see that $\eta_{\bar A\bar B}$ and $h^{\bar A}_B$ 
are the independent degrees of freedom. Also $h^{\bar A}_B$ is
a continuous (differentiable) variable while $\eta_{\bar A\bar B}$ is
a discrete (non-differentiable) variable. Thus the dynamics of
the metric signature, $\eta_{\bar A\bar B}$, can not be described 
by differential equations; one should apply a 
quantum description for these degrees of freedom. This description
could be stochastic in agreement with 't~Hooft's proposition 
that the origin of quantum gravity should be stochastic 
\cite{hooft99}. 
\par
In this case the basic question is: what kind of weight function
should be associated with each mode
($\eta_{\bar A\bar B} = \pm 1$ in our case). We will assume
\cite{dzh99nd}, \cite{dzh4} that this weight is connected with
the algorithmic complexity (AC) of a given mode. The notion of 
AC was first introduced by Kolmogorov \cite{kol1} and
leads to an algorithmic understanding of probability. The
idea is simple: the probability for an object is connected 
with the \textit{minimal} length of an algorithm describing 
this object.  Kolmogorov showed how this definition could
be used to define a notion of probability. Such a definition of
probability \textbf{can be applied to a single object} and as such
is of great interest for quantum gravity. 
\par 
In this paper we expand the 5D model of \cite{dzh99nd} 
to 7D with an SU(2) gauge group.  
We consider a 7D Universe with a fluctuating metric 
signature ($\eta_{\bar 0\bar 0} = \pm1$) and show that from 
the 4D point of view we obtain an SU(2) instanton field configuration 
and frozen extra dimensions (ED). The SU(2) solution given here
in the context of higher dimensional gravity is related to
the SU(2) \cite{hoso} \cite{don} and higher gauge groups \cite{bert}
wormhole instantons solutions to the coupled Einstein-Yang-Mills
(EYM) systems in 4D.
\par
Refs. \cite{hoso} \cite{bert} examined the cosmological
consequences of these 4D EYM wormhole solutions. For
example, Hosoya and Ogura \cite{hoso} related their
solution to a wormhole tunneling amplitude and the
Coleman mechanism \cite{cole} for the vanishing of the cosmological
constant. In this paper we will also investigate the
cosmological consequences of our solution. We will
argue that a small piece of our solution, with the linear size
$\approx l_{Pl}$, can be interpreted as giving rise to
the quantum birth of a Universe as a result of the
fluctuating metric signature. Then
the evolution of an ordinary Lorentzian Universe can begin 
from a boundary of this $\approx l_{Pl}$ sized piece. Simultaneously
with the formation of this Lorentzian Universe
the ED \textit{split off}, \textit{i.e.} the $h^{\bar a}_b$
components become non-dynamical variables. Once
the ED become ``frozen'' the Lagrangian effectively
reduces to ordinary 4D Einstein-Yang-Mills gravity.
This scenario can be seen as the non-singular, quantum
birth of a Universe from ``Nothing'' which results
from fluctuations of the metric signature
at the Planck scale.

\section{Field equations}

The total space $E$ of the principal bundle with structural
group $\cal G$ can be taken as the space $E$ which is acted
on by $\cal G$. This group action determines
the factor-space ${\cal H}=E/{\cal G}$ (the base of the
principal bundle) with the 4D metric
\begin{equation}
ds^2_{(4)} = \eta_{\bar \mu\bar \nu} 
\left ( 
h^{\bar \mu}_\alpha dx^\alpha 
\right )
\left ( 
h^{\bar \nu}_\beta dx^\beta 
\right )
\label{b1}
\end{equation}
which is the last term in Eq. (\ref{int2}). This 
allows us to insert a 4D term in the MD action
\begin{eqnarray}
S = && \int \left (R + 2\Lambda_1 \right )
\sqrt{|G|}d^{4+N}x + 
\int 
\left (2\Lambda_2\right )
\sqrt{|g|}d^4x 
\nonumber \\
&=& \int 
\left [ 
\int \left (R + 2\Lambda_1 \right )
\sqrt{|\gamma|}d^Ny + 2\Lambda_2
\right ]
\sqrt{|g|}d^4x
\label{b2}
\end{eqnarray}
$R$ is the Ricci scalar and
$G_{AB} = \eta_{\bar C\bar D}h^{\bar C}_Ah^{\bar D}_B$ 
is the MD metric on the total space; 
$g_{\mu\nu} = \eta_{\bar\alpha\bar\beta}h^{\bar\alpha}_\mu
h^{\bar\beta}_\nu$ is the 4D metric on the base of the principal 
bundle;  $\gamma_{ab} = \eta_{\bar c\bar d}h^{\bar c}_ah^{\bar d}_b$ 
is the metric on $\cal G$; $G, g$ and 
$\gamma$ are the appropriate metric determinates; 
$\Lambda_{1,2}$ are the MD and 4D $\lambda$-constants; 
$N=dim({\cal G})$. The MD action of Eq. (\ref{b2}) has several
points in common with the 4D EYM action considered in
Ref. \cite{hoso} (non-zero cosmological constants and
effective SU(2) ``Yang-Mills'' gauge fields).  Eq. (\ref{b2})
also has a connection to the action for the Non-gravitating Vacuum
Energy Theory \cite {gund}. In Ref. \cite{gund} Guendelman considers
an action which has degrees of freedom which are independent of the
metric, with the resulting action having two measures of integration
(involving metric and non-metric degrees of freedom). Eq. (\ref{b2})
incorporates two distinct degrees of freedom : the continuous variables,
$h^{\bar A}_B$, and the discrete variables, $\eta_{\bar A\bar B}$.
In Ref. \cite{gund} both the metric and non-metric degrees of
freedom were continuous.
\par 
This choice of the action indicates that we 
restrict our coordinate transformation law to
\begin{eqnarray}
y'^a & = & y'^a \left (y^b\right ) + 
f^a\left ( x^\alpha \right ) ,
\label{b3}\\
x'^\mu & = & x'^\mu \left (x^\alpha \right ) .
\label{b4}
\end{eqnarray}
These coordinate transformations do not 
destroy the $\cal G$-structure of the total space of the 
principal bundle, {\it i.e.} they do not mix different fibres of 
the bundle.
\par 
The independent, \textit{continuous degrees} of freedom  
are: the vier-bein $h^{\bar\mu}_\nu(x^\alpha)$, 
the gauge potential $h^{\bar a}_\mu(x^\alpha)$ and the scalar field 
$b(x^\alpha)$ which is defined as
\begin{equation}
h^{\bar a}_b(x^\mu) = \sqrt{b(x^\mu)} e^{\bar a}_b 
\label{b5}
\end{equation}
$e^{\bar a}_b$ is defined as 
\begin{equation}
\omega^{\bar a} = e^{\bar a}_b dx^b
\label{b6}
\end{equation}
$x^b$ are the coordinates on the group $\cal G$; 
$\omega^{\bar a}$ are the 1-forms satisfying
\begin{equation}
d\omega^{\bar a} = f^{\bar a}_{\bar b\bar c}
\omega^{\bar b}\wedge\omega^{\bar c}
\label{b7}
\end{equation}
$f^{\bar a}_{\bar b\bar c}$ are the structural constants 
of $\cal G$. Varying the action in Eq. (\ref{b2}) with respect to
$h^{\bar\mu}_\nu$, $h^{\bar a}_\nu$ and
$b$ leads to (see the Appendix for details) 
\begin{eqnarray}
R_{\bar\mu\bar\nu} - \frac{1}{2}\eta_{\bar\mu\bar\nu} R & = & 
\eta_{\bar\mu\bar\nu}
\left (\Lambda_1 + \frac{\Lambda_2}{b^{3/2}} \right ) ,
\label{b8}\\
R_{\bar a\bar\mu} & = & 0 ,
\label{b9}\\
R^{\bar a}_{\bar a} & = & - \frac{6}{5}
\left (\Lambda_1 + \frac{\Lambda_2}{b^{3/2}} \right ) .
\label{b10}
\end{eqnarray}
Eq. (\ref{b8}) are the Einstein vacuum 
equations with $\lambda$-terms; Eq. (\ref{b9}) 
are the ``Yang-Mills'' equations; Eq. (\ref{b10}) 
is reminiscent of Brans-Dicke theory since
the metric on each fibre is symmetric and has only one 
degree of freedom - the scalar factor $b(x^\mu)$ defined in 
Eq. (\ref{b5}). 
\par
We now investigate Eqs. (\ref{b8})-(\ref{b10})
using the ansatz
\begin{equation}
ds^2 = \sigma dt^2 + b(t)
\left (\omega^{\bar a} + A^{\bar a}_\mu dx^\mu\right )
\left (\omega_{\bar a} + A_{\bar a\mu} dx^\mu\right ) + 
a(t)d\Omega^2_3
\label{b11}
\end{equation}
$\sigma=\pm1$ describes the possible quantum fluctuation 
of the metric signature between Euclidean and Lorentzian modes, 
$A^{\bar a}_\mu$ are SU(2) gauge potentials, 
$d\Omega^2_3=d\chi^2 + \sin^2\chi\left (
d\theta^2 + \sin^2\theta d\phi^2 \right )$ is the metric on the 
unit $S^3$ sphere and $x^0=t, x^1=\chi ,x^2=\theta ,
x^3=\phi , x^5=\alpha ,x^6=\beta ,x^7=\gamma$. 
($\alpha ,\beta ,\gamma$ are the Euler angles for the SU(2) 
group) 
\begin{eqnarray}
\omega^1 & = & {1\over 2}
(\sin \alpha d\beta - \sin \beta \cos \alpha d\gamma ),
\label{b12}\\
\omega^2 & = & -{1\over 2}(\cos \alpha d\beta +
\sin \beta \sin \alpha d\gamma ),
\label{b13}\\
\omega^3 & = & {1\over 2}(d\alpha +\cos \beta d\gamma ).
\label{b14}
\end{eqnarray}
The nondiagonal components of the MD metric 
take the instanton-like form \cite{don} \cite{instanton} : 
\begin{eqnarray}
A^a_\chi & = & \frac{1}{4}\left \{ -\sin\theta \cos\varphi ;
-\sin\theta \sin\varphi ;\cos\theta \right \}
(f(t) - 1),
\label{b16}\\
A^a_\theta & = & \frac{1}{4}\left \{ -\sin\varphi ;
-\cos\varphi ;0\right \}(f(t) - 1),
\label{b17}\\
A^a_\varphi & = & \frac{1}{4}\left \{0;0;1\right \}
(f(t) - 1).
\label{b18}
\end{eqnarray}
Substituting into Eqs. (\ref{b8})-(\ref{b10}) gives 
\begin{eqnarray}
\frac{1}{3}R^{\bar a}_{\bar a} = R^{\bar 5}_{\bar 5} = 
-\frac{\sigma}{2}\frac{\ddot b}{b} + \frac{2}{b} - 
\frac{\sigma}{4}\frac{{\dot b}^2}{b^2} - 
\frac{3}{4}\sigma \frac{\dot a\dot b}{ab} + 
\frac{1}{8}\frac{b}{a}
\left (\sigma E^2 + H^2 \right )
& = & -\frac{2}{5}\left (\Lambda_1 + 
\frac{2\Lambda_2}{b^{3/2}} \right ) ,
\label{b19}\\
G_{\bar 0\bar 0} = 
-3\frac{\sigma}{b} + \frac{3}{4}\frac{{\dot b}^2}{b^2} - 
3\frac{\sigma}{a} + 
\frac{9}{4}\frac{\dot a\dot b}{ab} + 
\frac{3}{16}\frac{{\dot a}^2}{a^2} - 
\frac{3}{16}\frac{b}{a}\left (E^2 - \sigma H^2 \right ) 
& = & \sigma \left (\Lambda_1 + 
\frac{\Lambda_2}{b^{3/2}} \right ) ,
\label{b20}\\
G_{\bar 1\bar 1} = \frac{3}{2}\sigma\frac{\ddot b}{b} - 
\frac{3}{b} + \sigma \frac{\ddot a}{a} - 
\frac{1}{a} + \frac{3}{2}\sigma\frac{\dot a\dot b}{ab} - 
\frac{\sigma}{4}\frac{{\dot a}^2}{a^2} + 
\frac{1}{16}\frac{b}{a}\left (\sigma E^2 - H^2 \right ) 
& = & \left (\Lambda_1 + 
\frac{\Lambda_2}{b^{3/2}} \right ) ,
\label{b21}\\
G_{\bar 2\bar 7} = 
2\ddot f + 5\frac{\dot b\dot f}{b} + 
\frac{\dot a\dot f}{a} - 
4\frac{\sigma}{a}f\left (f^2 - 1 \right ) 
& = & 0 ,
\label{b22}\\
E^2 = E^a_i E^{ai} = \dot f^2 , 
\quad 
H^2 = H^a_i H^{ai} & = & \frac{
\left (f^2 - 1 \right )^2}{a} ,
\label{b23}
\end{eqnarray}
$G_{\bar A\bar B} = R_{\bar A\bar B} - 
(1/2)\eta_{\bar A\bar B} R$; $i=1,2,3$ are space indices;
the ``electromagnetic'' fields are
\begin{equation}
E^a_i = F^a_{0i}, \quad H^a_i = 
\frac{1}{2}\varepsilon_{ijk} F^{ajk}
\label{b24}
\end{equation}
$F^a_{\mu\nu}$ is the field strength tensor for the non-Abelian
gauge group. The wormhole instanton of Ref. \cite{hoso}
had a vanishing ``electric'' field. In contrast the
solution studied here has both non-vanishing ``electric''
and ``magnetic'' fields.

\section{Definition of algorithmic complexity}

We now examine the dynamic behavior of the \textit{discrete}  
quantity $\sigma$ which describes the quantum fluctuations 
(trembling) between Euclidean and Lorentzian 
modes. One fruitful approach is  
the stochastic approach proposed by 't~Hooft 
\cite{hooft99}. The main question is: how
to define a weight function for each mode (Euclidean and
Lorentzian) ? Our proposition is that these weight functions
be given by the AC of the Eqs. (\ref{b19})-(\ref{b22}). 
\par 
In detail Eqs. (\ref{b19})-(\ref{b22}) 
define the dynamic behavior of the continuous
variables $a(t), b(t)$ and $f(t)$.
Each equation oscillates between the two possibilities
$\sigma = \pm 1$, and when viewed as an algorithm,
will have an AC which depends on the value of $\sigma$.
Based on the AC, each equation is assigned two weight functions:
one for $\sigma = +1$ and one for $\sigma = -1$.
Certain equations will be simpler in the Euclidean mode
while others will be simpler in the Lorentzian mode. A
common example of this behavior is the Polyakov-`t~Hooft instanton
which exists only in Euclidean space. 
\par 
Kolmogorov's \cite{kol1} definition for AC is :
\par
\textit{
The algorithmic complexity $K(x\mid y)$ of the  object $x$  for a 
given object $y$ is the minimal length of the ``program'' $P$
that is written as a sequence of  the  zeros  and  unities
which allows us to construct $x$ having $y$:
\begin{equation}
K(x\mid y) = \min_{A( P,y)=x} l(P)
\label{ac1}
\end{equation}
where $l(P)$ is length of the  program $P$; $A(P,y)$  is  the
algorithm  for calculating an object $x$, using  the  program $P$,
when the object $y$ is given.
} 
\par
This definition gives an exact mathematical meaning to
the word ``simple''. It is also in the spirit of
Einstein's statement: 
\textit{``Everything should be simple as possible but not 
more''}.  

\section{Quantum fluctuation for the initial equations}

Our assumption of quantum trembling 
between Euclidean and Lorentzian modes is described by as a 
quantum-stochastic fluctuation between the equations
\begin{equation}
\begin{array}{ccc}
\sigma = +1 & \longleftrightarrow & \sigma = -1
\\
& \Downarrow  & 
\\
\left (R^+\right )^{\bar 5}_{\bar 5} 
& 
\longleftrightarrow
& 
\left (R^-\right )^{\bar 5}_{\bar 5} 
\\
\left (G^+\right )_{\bar 0\bar 0} 
& 
\longleftrightarrow
& 
\left (G^-\right )_{\bar 0\bar 0}
\\
\left (G^+\right )_{\bar 1\bar 1} 
& 
\longleftrightarrow
& 
\left (G^-\right )_{\bar 1\bar 1}
\\
\left (G^+\right )_{\bar 2\bar 7} 
& 
\longleftrightarrow
& 
\left (G^-\right )_{\bar 2\bar 7}
\end{array}
\label{qf1}
\end{equation}
The signs ($\pm$) denote the equations of 
the Euclidean (+) or Lorentzian (-) mode. Now we 
define the weight functions for each pair in Eqs. (\ref{qf1}). 

\subsection{$G_{\bar 2\bar 7}$ equation}

This equation in the Euclidean mode is 
\begin{equation}
2\ddot f + 5\frac{\dot b\dot f}{b} + 
\frac{\dot a\dot f}{a} - 
\frac{4}{a}f\left (f^2 - 1 \right )  = 0 
\label{in1}
\end{equation}
which has the instanton solution 
\begin{equation}
\dot f = \frac{1 - f^2}{\sqrt a} ,
\label{in2}
\end{equation}
where
\begin{equation}
b = b_0 = const
\label{in3}
\end{equation}
Eq. (\ref{in2}) implies the instanton condition 
\begin{equation}
E^a_i E_a^i = H^a_i H_a^i .
\label{in4}
\end{equation}
In the Lorentzian mode
\begin{equation}
2\ddot f + 5\frac{\dot b\dot f}{b} + 
\frac{\dot a\dot f}{a} + 
\frac{4}{a}f\left (f^2 - 1 \right )  = 0 
\label{in6}
\end{equation}
and the instanton solution (\ref{in4}) is not a 
solution of (\ref{in6}). It is well known 
that the non-singular, instanton solution 
exists only in Euclidean space.
\par
In terms of the AC criteria the Euclidean equation (\ref{in1}) 
is simpler than Lorentzian equation (\ref{in6}), since it
is equivalent to the first order differential equation (\ref{in2}). 
\par
In a first rough approximation we set
the probability of the $G_{\bar 2\bar 7}=0$ equation for
the Euclidean mode to $p^+_{27}=1$ and the 
Lorentzian mode to $p^-_{27}=0$. 
\par
The exact definition for 
each $p^\pm_{AB}$ probability is \cite{dzh4} 
\begin{equation}
p^\pm_{AB} = \frac{e^{-K^\pm_{AB}}}
{e^{-K^+_{AB}} + e^{-K^-_{AB}}}
\label{in7}
\end{equation}
where $K^\pm_{AB}$ is the AC for the appropriate 
equation. If $K^+_{27} \ll K^-_{27}$ we have 
$p^+_{27} = 1$ and $p^-_{27} = 0$. The expression
for the probability in Eq. (\ref{in7}) can be
seen as the discrete variable analog of the Euclidean path
integral transition probability. For a continuous variable the
Euclidean path integral gives the probability for the variable
to evolve from some initial configuration to some
final configuration as being proportional to the
exponential of minus the action ($\propto e^{-S}$).
Eq. (\ref{in7}) is similar, but with the AC replacing
the action. The denominator normalizes the probability
(it is a sum rather than integral since we are dealing
with a discrete variable).

\subsection{$R^{\bar 5}_{\bar 5}$ equation}

This equation in the Euclidean mode is 
\begin{equation}
-\frac{1}{2}\frac{\ddot b}{b} + \frac{2}{b} - 
\frac{1}{4}\frac{{\dot b}^2}{b^2} - 
\frac{3}{4}\frac{\dot a\dot b}{ab} + 
\frac{1}{8}\frac{b}{a}
\left (E^2 + H^2 \right ) = 
-\frac{2}{5}\left (\Lambda_1 + 
\frac{2\Lambda_2}{b^{3/2}} \right ) ,
\label{a1}
\end{equation}
and in the Lorentzian mode 
\begin{equation}
\frac{1}{2}\frac{\ddot b}{b} + \frac{2}{b} + 
\frac{1}{4}\frac{{\dot b}^2}{b^2} + 
\frac{3}{4}\frac{\dot a\dot b}{ab} + 
\frac{1}{8}\frac{b}{a}
\left (-E^2 + H^2 \right ) = 
-\frac{2}{5}\left (\Lambda_1 + 
\frac{2\Lambda_2}{b^{3/2}} \right ) ,
\label{a2}
\end{equation}
The Lorentzian mode equation is simpler because the two last terms 
annihilate as a consequence of the instanton condition 
(\ref{in4}). 
\par
To a first rough approximation 
we set the probability of the $R^{\bar 5}_{\bar 5}$ equation for 
the Euclidean mode to $p^+_{55}=0$ and the 
Lorentzian mode to $p^-_{55}=1$. 

\subsection{$G_{\bar 0\bar 0}$ equation}

This equation in the Euclidean mode is 
\begin{equation}
-\frac{3}{b} + \frac{3}{4}\frac{{\dot b}^2}{b^2} - 
\frac{3}{a} + 
\frac{9}{4}\frac{\dot a\dot b}{ab} + 
\frac{3}{16}\frac{{\dot a}^2}{a^2} - 
\frac{3}{16}\frac{b}{a}\left (E^2 - H^2 \right ) = 
\left (\Lambda_1 + \frac{\Lambda_2}{b^{3/2}} \right ) 
\label{tt1}
\end{equation}
and in the Lorentzian mode 
\begin{equation}
\frac{3}{b} + \frac{3}{4}\frac{{\dot b}^2}{b^2} + 
\frac{3}{a} + 
\frac{9}{4}\frac{\dot a\dot b}{ab} + 
\frac{3}{16}\frac{{\dot a}^2}{a^2} - 
\frac{3}{16}\frac{b}{a}\left (E^2 + H^2\right ) = 
-\left (\Lambda_1 + \frac{\Lambda_2}{b^{3/2}} \right ) .
\label{tt2}
\end{equation}
In this case because of the instanton condition 
(\ref{in4}) the Euclidean equation is simpler and 
therefore in the first rough approximation we can 
set the probability of the $G_{\bar 0\bar 0}=0$ equation for 
the Euclidean mode to $p^+_{00}=1$ and the 
Lorentzian mode to $p^-_{00}=0$. 

\subsection{$G_{\bar 1\bar 1}$ equation}

This equation in the Euclidean mode is 
\begin{equation}
\frac{3}{2}\frac{\ddot b}{b} - 
\frac{3}{b} + \frac{\ddot a}{a} - 
\frac{1}{a} + \frac{3}{2}\frac{\dot a\dot b}{ab} - 
\frac{1}{4}\frac{{\dot a}^2}{a^2} + 
\frac{1}{16}\frac{b}{a}\left (E^2 - H^2 \right ) = 
\left (\Lambda_1 + \frac{\Lambda_2}{b^{3/2}} \right ) 
\label{cc1}
\end{equation}
and in the Lorentzian mode 
\begin{equation}
-\frac{3}{2}\frac{\ddot b}{b} - 
\frac{3}{b} - \frac{\ddot a}{a} - 
\frac{1}{a} - \frac{3}{2}\frac{\dot a\dot b}{ab} + 
\frac{1}{4}\frac{{\dot a}^2}{a^2} - 
\frac{1}{16}\frac{b}{a}\left (E^2 + H^2 \right ) = 
\left (\Lambda_1 + \frac{\Lambda_2}{b^{3/2}} \right ) .
\label{cc2}
\end{equation}
As in the previous subsection as a 
consequence of the instanton condition (\ref{in4}) the 
Euclidean mode is simpler. 
Therefore in the first rough approximation we set 
$p^+_{11}=1$ and $p^-_{11}=0$. 

\subsection{Mixed system of equations}

The mixed system of equations for the 7D spacetime 
with fluctuating metric signature is 
\begin{eqnarray}
2\ddot f + 5\frac{\dot b\dot f}{b} + 
\frac{\dot a\dot f}{a} - 
\frac{4}{a}f\left (f^2 - 1 \right ) & = & 0 ,
\label{m1}\\
\frac{1}{2}\frac{\ddot b}{b} + \frac{2}{b} + 
\frac{1}{4}\frac{{\dot b}^2}{b^2} + 
\frac{3}{4}\frac{\dot a\dot b}{ab} + 
\frac{1}{8}\frac{b}{a}
\left (-E^2 + H^2 \right ) & = & 
-\frac{2}{5}\left (\Lambda_1 + 
\frac{2\Lambda_2}{b^{3/2}} \right ) ,
\label{m2}\\
-\frac{3}{b} + \frac{3}{4}\frac{{\dot b}^2}{b^2} - 
\frac{3}{a} + 
\frac{9}{4}\frac{\dot a\dot b}{ab} + 
\frac{3}{16}\frac{{\dot a}^2}{a^2} - 
\frac{3}{16}\frac{b}{a}\left (E^2 - H^2 \right ) & = & 
\left (\Lambda_1 + \frac{\Lambda_2}{b^{3/2}} \right ) ,
\label{m3}\\
\frac{3}{2}\frac{\ddot b}{b} - 
\frac{3}{b} + \frac{\ddot a}{a} - 
\frac{1}{a} + \frac{3}{2}\frac{\dot a\dot b}{ab} - 
\frac{1}{4}\frac{{\dot a}^2}{a^2} + 
\frac{1}{16}\frac{b}{a}\left (E^2 - H^2 \right ) & = & 
\left (\Lambda_1 + \frac{\Lambda_2}{b^{3/2}} \right )  .
\label{m4}
\end{eqnarray}
The solution for this system is
\begin{eqnarray}
a & = & t^2 ,
\label{m5}\\
f & = & \frac{t^2 - t_0^2}{t^2 + t_0^2} ,
\label{m6}\\
b & = & b_0 = const ,
\label{m7}\\
\Lambda_1 & = & -\frac{1}{b_0} ,
\label{m8}\\
\Lambda_2 & = & -2\sqrt b_0 .
\label{m9}
\end{eqnarray}
The existence of this solution is somewhat surprising ! 
Let us clarify this. Normally in any dimension the Bianchi identities
are fulfilled. Therefore some gravitational field
equations are not independent of the others.
Ordinarily the superfluous equations are associated with
initial conditions (\textit{i.e.} Eq. (\ref{m3}) above). 
In our case the mixed system above comes from a
model with a varying metric signature. As a
consequence the  Bianchi identities are not correct and 
this system should be unsolvable. Evidently the 
solution is a condition for the solvability of the mixed system 
which uniquely define the $\Lambda$-constants. 
If the solution in Eqs. (\ref{m5})-(\ref{m9}) is unique then it
must be absolutely stable. 
\par 
The physical meaning of this solution is: 
\begin{itemize}
\item
Eq. (\ref{m5}) implies a flat 4D Einstein spacetime 
that is not effected by matter.
\item
Eq. (\ref{m6}) implies a Polyakov - 't Hooft instanton
gauge field configuration which is not effected by gravity.
\item 
Eq. (\ref{m7}) implies a frozen ED. 
\item
Eqs. (\ref{m8})-(\ref{m9}) imply that the dynamical 
equations uniquely determine the $\Lambda_{1,2}$-constants.
\end{itemize}
It is interesting to note that the effective cosmological
constant terms on the right hand side of Eqs. (\ref{b8})
(\ref{b10}) ({\it i.e.} $\Lambda_1$ and $\Lambda_2 /b^{3/2}$)
are inversely proportional to the size of the ED, $b_0$. Thus
in order to have a small cosmological constant term one needs
to have a large ED. This could be seen as supporting 
the large extra dimensions scenarios \cite{hamed}.

\section{Physical applications of the solution}

\subsection{Regular Universe}

We can interpret this solution as a flat 
4D Universe with fluctuating metric signature, filled 
with an SU(2) instanton gauge field and frozen ED. Astonishingly 
this Universe has only one manifestation of gravity: the 
frozen ED that result from the fluctuating metric 
signature. This model Universe is a simple example of possible
effects connected with the dynamics of non-differentiable
variables.

\subsection{Non-singular birth of the Universe}

Various researchers (see Ref. \cite{hawking} for example)
have speculated about the quantum birth
of the Universe from ``Nothing''. In light of this we can
interpret a small piece (with linear size $\approx l_{Pl}$) of
our model 7D Universe as a quantum birth of the regular 
4D Universe. In contrast to other scenarios 
this origin has a metric signature trembling between 
Euclidean and Lorentzian modes. Further we postulate that
on a boundary of this origin there occurs  
\begin{itemize}
\item
\textit{a quantum transition} to only one Lorentzian mode 
of fixed metric signature.
\item
\textit{a splitting off} the ED so that the metric on the 
fibres $(h^{\bar a}_b)$ becomes a non-dynamical variable. 
After this splitting off the linear size of the gauge group remains 
constant yielding ordinary 4D Einstein-Yang-Mills 
gravity. 
\end{itemize}
These assumptions about a quantum transition from 
fluctuating metric signature $(\pm 1,+1, \cdots ,+1)$ 
to Lorentzian signature $(-1,+1, \cdots ,+1)$ and
a splitting off of the ED should not be seen as something
extraordinary and new, but rather as an extension of our
postulate about the quantum birth of the regular 4D
Universe, discussed above, with certain laws
(gravitational equations + non-differentiable
dynamic). The present case can be seen a quantum-stochastic  
change or
evolution of these laws (here this involves only the quantum
transition of $\eta_{00}$ and the splitting off of the ED). 
\par
The probability for the quantum birth is 
\begin{equation}
P \approx N e^{-S}
\label{ns1}
\end{equation}
where $S$ is the Euclideanized, dimensionless action,
which should be $S \approx 1$ in Planck units.
The prefactor $N$ is of more interest,
since it contains information about the topological
structure of the boundary of the origin. 
\par 
The probability for the quantum-stochastic transition to Lorentzian 
mode and splitting off of the ED should be determined by 
the AC of the final and initial states. Such a quantum-stochastic 
transition 
can occur only if the final state with Lorentzian mode and splitting 
off the ED is simpler than the initial state with the fluctuating 
metric signature and dynamic ED.

\section{Conclusions}

In this article we have investigated possible
quantum gravity effects connected with
\textit{non-differentiable degrees of freedom} \cite{dzh99nd},
\cite{dzh4}. By considered the quantum trembling
of the metric signature of a 7D model Universe
we have found a solution which describes 
\textit{a flat 
4D Universe with a fluctuating metric signature filled with an 
SU(2) instanton gauge field and frozen ED}. A piece of this 
solution can be considered \textit{as
resulting in the quantum birth of the regular
Universe} with the fluctuating metric signature. 
An important peculiarity for this model is that 
it is a \textit{vacuum} model without any kind of 
matter; only the gauge field appear as non-diagonal 
components of the MD metric. This is in the spirit of Einstein's 
point of view that Nature consists of ``Nothing''.

\section{Appendix}

We start from the Lagrangian adopted for the vacuum gravitational 
theory on the principal bundle with the structural group 
$\cal G$ ($\dim ({\cal G}) = N$). $\cal G$ is the gauge group
associated with the EDs
\begin{equation}
S = \int \left (R + 2\Lambda_1 \right )
\sqrt{|G|}d^{4+N}x + 
\int 
\left (2\Lambda_2'\right )
\sqrt{|g|}d^4x 
\label{ap1}
\end{equation}
where $R$ is the Ricci scalar for the total space; $G$ 
and $g$ are the determinant of the metric on the total 
space and base of the principal bundle respectively, 
$\Lambda_1, \Lambda_2'$ are the MD and 4D $\lambda$-constants. This 
Lagrangian is correct if the coordinate transformations 
conserve the topological structure of the total space 
({\it i.e.} does not mix the fibres)
\begin{eqnarray}
y'^a & = & y'^a \left (y^b\right ) + 
f^a\left ( x^\alpha \right ) ,
\label{ap2}\\
x'^\mu & = & x'^\mu \left (x^\alpha \right ) .
\label{ap3}
\end{eqnarray}
The metric on the total space can be written as
\begin{eqnarray}
ds^2_{(MD)} & = & 
\left ( 
\sqrt{b}\omega^{\bar a} + h^{\bar a}_\mu dx^\mu
\right )
\left ( 
\sqrt b\omega_{\bar a} + h_{\bar a\mu} dx^\mu
\right ) + 
\left ( 
h^{\bar \mu}_\alpha dx^\alpha 
\right )
\left ( 
h_{\bar \mu\beta} dx^\beta 
\right )
\label{ap4}\\
\omega^{\bar a} & = & e^{\bar a}_b dx^b 
\quad
h^{\bar a}_b = \sqrt{b(x^\mu)} e^{\bar a}_b
\label{ap5}
\end{eqnarray}
where $x^\mu$ and $y^b$ are the coordinates 
along the base and fibres respectively; 
(Greek indices)$=0,1,2,3$ and (Latin indices)$=5,6,\cdots ,N$; 
$\bar A = \bar a,\bar \mu$ is the viel-bein index; 
$\eta_{\bar A\bar B} = \{\pm 1,\pm 1, \cdots ,\pm 1\}$
is the signature of the MD metric; $\omega^{\bar a}$ are 
the 1-forms satisfying to the structural equations 
\begin{equation}
d\omega^{\bar a} = f^{\bar a}_{\bar b\bar c}
\omega^{\bar b}\wedge\omega^{\bar c}
\label{ap6}
\end{equation}
where $f^{\bar a}_{\bar b\bar c}$ are the structural 
constants for the gauge group $\cal G$.
\par
The independent degrees of freedom for gravity on the 
principal bundle with the structural group ${\cal G}$ is 
vier-bein $h^{\bar\mu}_\nu(x^\alpha)$, gauge potential 
$h^{\bar a}_\nu(x^\alpha)$ and scalar field $b(x^\alpha)$
\cite{sal1,per1,coq1983zn}. 
All functions depend only on the point $x^\mu$ on the base 
of the principal bundle as a consequence of the symmetry of the
fibres.
\par
Varying the action (\ref{ap1}) with respect to 
$h^{\bar\mu}_\nu(x^\alpha)$ leads to 
\begin{equation}
\int\left (
R^\mu_{\bar\nu} - \frac{1}{2}h^\mu_{\bar\nu} R - 
\Lambda_1h^\mu_{\bar\nu} 
\right )
\sqrt {|\gamma|} d^Ny - \Lambda_2' h^\mu_{\bar\nu} = 0
\label{ap7}
\end{equation}
where $|\gamma| = \det h^{\bar a}_b = b^N\det e^{\bar a}_b$ 
is the volume element on the fibre and 
$\sqrt {|G|} = \sqrt {|g|}\sqrt {|\gamma|}$ is a consequence of 
the following structure of the MD metric
\begin{eqnarray}
h & = & h^{\bar A}_B =  
\left (
\begin{array}{cc}
h^{\bar a}_b & h^{\bar a}_\mu \\
0 & h^{\bar \nu}_\mu
\end{array}
\right ) ,
\label{ap8} \\
h^{-1} & = & h^B_{\bar A} = 
\left (
\begin{array}{cc}
h^b_{\bar a} & -h^b_{\bar a}h^{\bar a}_\nu h^\nu_{\bar\nu} \\
0 & h^\mu_{\bar \nu}
\end{array}
\right ) ,  
\label{ap9} \\
h^b_{\bar a} & = & {\left (h^{\bar a}_b \right )} 
^{-1}
\quad
h^\mu_{\bar \nu} = {\left (h^{\bar \nu}_\mu \right )} 
^{-1} .
\label{ap10}
\end{eqnarray}
An integration over the EDs can be easily performed since
no functions depend on $y^a$ 
\begin{equation}
\int \left (\cdots\right ) \sqrt {|\gamma|} d^Ny = 
\left (\cdots\right ) \int \sqrt {|\gamma|} d^Ny = 
\left (\cdots\right ) b^{N/2} V_{\cal G}
\label{ap11}
\end{equation}
where ${V_{\cal G}} = \int \sqrt{\det (e^{\bar a}_b)} d^Ny$  
is the volume of the gauge group $\cal G$. 
In this case Eq. (\ref{ap7}) becomes
\begin{equation}
R^\mu_{\bar\nu} - \frac{1}{2}h^\mu_{\bar\nu} R = 
\left (
\Lambda_1 + \frac{\Lambda_2}{b^{N/2}}
\right )
h^\mu_{\bar\nu}
\label{ap12}
\end{equation}
where $\Lambda_2' = V_{\cal G}\Lambda_2$. 
\par
Varying with respect to $h^{\bar a}_\mu(x^\alpha)$ leads to 
\begin{equation}
R^\mu_{\bar a} = 0
\label{ap13}
\end{equation}
as $h^{\bar a}_\mu$ does not consists in 
$\det (h^{\bar A}_B) = \det (h^{\bar a}_b) \det (h^{\bar\mu}_\nu)$. 
\par
Varying with respect to $b(x^\alpha)$ leads to
\begin{equation}
\frac{\delta S}{\delta b} = \sum_{\bar a,b} 
\frac{\delta h^{\bar a}_b}{\delta b} \frac{\delta S}{\delta h^{\bar a}_b} 
= h^{\bar a}_A \left (
R^A_{\bar a} - \frac{1}{2} h^A_{\bar a} 
- \Lambda_1 h^A_{\bar a}
\right )
\label{ap14}
\end{equation}
here we used Eq. (\ref{ap12}) and $h^\mu_{\bar a} = $. This 
equation we write in the form
\begin{equation}
R^{\bar a}_{\bar a} - \frac{N}{2} R = 
N \Lambda_1
\label{ap15}
\end{equation}
From Eq. (\ref{ap12}) we have 
\begin{eqnarray}
h^{\bar\nu}_\mu
\left [
R^\mu_{\bar\nu} - \frac{1}{2} h^\mu_{\bar\nu} R - 
\left (
\Lambda_1 + \frac{\Lambda_2}{b^{N/2}}
\right )h^\mu_{\bar\nu} R
\right ] = 
h^{\bar\nu}_\mu \left [ \cdots \right ] + 
h^{\bar\nu}_a \left [ \cdots \right ] & = & 
\nonumber \\
h^{\bar\nu}_A
\left [
R^A_{\bar\nu} - \frac{1}{2} h^A_{\bar\nu} R - 
\left (
\Lambda_1 + \frac{\Lambda_2}{b^{N/2}}
\right )h^A_{\bar\nu} R
\right ] = 
R^{\bar\nu}_{\bar\nu} - 2 R - 4
 \left (
\Lambda_1 + \frac{\Lambda_2}{b^{N/2}}
\right ) & = & 0
\label{ap16}
\end{eqnarray}
Adding Eqs. (\ref{ap16}) and (\ref{ap15}) we find
\begin{equation}
R = R^{\bar A}_{\bar A} = - \frac{2}{N+2}
\left [
\left ( N+4\right )\Lambda_1 + 
\frac{4\Lambda_2}{b^{N/2}}
\right ]
\label{ap17}
\end{equation}
Finally we have
\begin{eqnarray}
R^{\bar a}_{\bar a} & = & -\frac{2N}{N+2} 
\left (
\Lambda_1 + \frac{\Lambda_2}{b^{N/2}}
\right ) ,
\label{ap18} \\
R^\mu_{\bar a} & = & 0
\label{ap19} \\
R^\mu_{\bar\nu} - \frac{1}{2}h^\mu_{\bar\nu} R & = & 
\left (
\Lambda_1 + \frac{\Lambda_2}{b^{N/2}}
\right )
h^\mu_{\bar\nu}
\label{ap20}
\end{eqnarray}
This equation system can be rewritten as 
\begin{eqnarray}
R^{\bar a}_{\bar a} & = & -\frac{2N}{N+2} 
\left (
\Lambda_1 + \frac{\Lambda_2}{b^{N/2}}
\right ) ,
\label{ap21} \\
R_{\bar\mu\bar a} & = & 0
\label{ap22} \\
R_{\bar\mu\bar\nu} - \frac{1}{2}\eta_{\bar\mu\bar\nu} R & = & 
\left (
\Lambda_1 + \frac{\Lambda_2}{b^{N/2}}
\right )
\eta_{\bar\mu\bar\nu}
\label{ap23}
\end{eqnarray}
here we have used $h^{\bar\nu}_b = 0$.

\section{Acknowledgments}
VD is supported by a Georg Forster Research Fellowship
from the Alexander von Humboldt Foundation and 
grateful H.-J. Schmidt
for invitation to Potsdam Universit\"at for research.

\bibliography{john,gravity}
\bibliographystyle{prsty}

\end{document}